\documentclass[twocolumn,prl,tightenlines,nofootinbib,superscriptaddress,showpacs]{revtex4}

\usepackage{amsmath}
\usepackage{amssymb,amsfonts,latexsym}
\usepackage{bm}
\usepackage[mathcal]{euscript}
\usepackage{graphicx,epstopdf}
\usepackage{epsfig}
\usepackage{color}
\usepackage{units}
\usepackage{xfrac}

\begin{document}

\title{Simple and faithful nonlinear field equations for aligning self-propelled rods}

\author{Anton Peshkov}
\affiliation{Service de Physique de l'Etat Condens\'e, CEA-Saclay, URA 2464 CNRS, 91191 Gif-sur-Yvette, France}
\affiliation{LPTMC, CNRS-UMR 7600, Universit\'e Pierre et Marie Curie, 75252 Paris, France}
\affiliation{Max Planck Institute for the Physics of Complex Systems, N\"othnitzer Str. 38, 01187 Dresden, Germany}

\author{Igor S. Aranson}
\affiliation{Materials Science Division, Argonne National Laboratory, 9700
  South Cass Avenue, Argonne, IL 60439}
\affiliation{Max Planck Institute for the Physics of Complex Systems, N\"othnitzer Str. 38, 01187 Dresden, Germany}

\author{Eric Bertin}
\affiliation{Universit\'e de Lyon, Laboratoire de Physique, ENS Lyon, CNRS, 46 All\'ee d'Italie, 69007 Lyon, France}
\affiliation{Max Planck Institute for the Physics of Complex Systems, N\"othnitzer Str. 38, 01187 Dresden, Germany}

\author{Hugues Chat\'{e}}
\affiliation{Service de Physique de l'Etat Condens\'e, CEA-Saclay, URA 2464 CNRS, 91191 Gif-sur-Yvette, France}
\affiliation{Max Planck Institute for the Physics of Complex Systems, N\"othnitzer Str. 38, 01187 Dresden, Germany}

\author{Francesco Ginelli}
\affiliation{Institute for Complex Systems and Mathematical Biology,
King's College, University of Aberdeen, Aberdeen AB24 3UE, United Kingdom}

\date{\today}
\pacs{05.65.+b, 45.70.Vn, 87.18.Gh}

\begin{abstract}
We derive a set of minimal yet complete nonlinear field equations
describing the collective properties of self-propelled rods from a
simple microscopic starting point, the Vicsek model with nematic alignment.
Analysis of their linear and nonlinear dynamics
shows good agreement
with the original microscopic model.
In particular, we derive an explicit
expression for the fronts forming density-segregated, banded solutions,
allowing us to develop a more complete analytic picture of the problem at the nonlinear level.
\end{abstract}

\maketitle

Collective motion is a central theme in the rapidly growing field of
active matter studies which loosely groups together all situations where
energy is spent locally to produce coherent motion \cite{SR-REVIEW}.
In spite of the emergence of better-controlled, larger-scale
experiments \cite{DYCOACT,Schaller,Swinney,Sokolov,Sumino2012}, our understanding of collective motion mostly
comes from the study of mathematical models, and particularly of models
of ``dry'' active matter systems,
where the fluid which surrounds the moving objects can be neglected.

Microscopic models then usually consist of interacting self-propelled particles,
as in the Vicsek model \cite{VICSEK}, where constant-speed point particles
``ferromagnetically''
align their velocities with that of local neighbors.
The study of these models, together with
some more theoretical approaches, revealed a wealth of phenomena
such as true long-range orientational order in two dimensions,
spontaneous segregation of dense/ordered regions, anomalously strong
number fluctuations, etc. \cite{TTR-review,GNF,CHATE}.

These results have given rise to an emerging picture of universality classes,
typically depending on the symmetries involved,
which one would ideally characterize by some coarse-grained field equations.
Different routes can be followed to obtain such equations: they can be written
a priori, putting down all terms allowed by symmetries up to some arbitrary order
in gradients once the hydrodynamic fields have been identified.
One can also derive continuous descriptions from some microscopic starting point
under more or less controlled and constraining assumptions,
yielding more or less complete, well-behaved equations.
There is nevertheless shared belief, based mostly on
renormalization-group approaches,
that in each case there exists a set of minimal equations
accounting for all large-scale physics.

For polar particles aligning ferromagnetically (as in the Vicsek model),
there is now near-consensus about this minimal set of equations:
the phenomenological theory, initially proposed by Toner and Tu \cite{TT} is essentially correct
if one takes into account the dependences of its coefficients on density and
parameters initially overlooked but later derived
from microscopics in \cite{BDG,IHLE}.
It has been shown to reproduce many of the
phenomena observed in microscopic models \cite{MCM-polar}, although
a complete study of its nonlinear solutions and dynamics is still lacking.

For the other important universality class of polar particles
aligning {\it nematically} ---{\it e.g.} self-propelled rods---, the situation is
less satisfactory: Baskaran and Marchetti \cite{MCM-rod1} first derived rather lengthy
yet mostly linear hydrodynamic equations for hard rods interacting via excluded volume,
showing in particular the presence of global nematic, not polar, order,
in agreement with microscopic observations \cite{RODS}. Very recently \cite{MCM-rod2},
they added some nonlinear terms and performed a linear stability analysis
of the homogeneous ordered state within an arbitrary choice of parameters.
These even longer equations do not benefit from the same consensus as
the Toner-Tu theory.

In this Letter,
we derive a set of minimal yet complete nonlinear field equations
describing the collective properties of self-propelled rods from a
simple microscopic starting point, the Vicsek model with nematic alignment
studied in \cite{RODS}.
We use a ``Boltzmann-Ginzburg-Landau'' approach,
a controlled expansion scheme \cite{IGOR-LEV} which is a refined version
of that used in \cite{BDG}.
Analysis of the linear and nonlinear dynamics
of these equations
shows good agreement
with the original microscopic model.
In particular, we derive explicit
expressions for the fronts forming density-segregated, banded solutions,
allowing us to develop a more complete analytic picture of the problem at the nonlinear level.

The Vicsek model with nematic alignment \cite{RODS} consists of point particles moving off-lattice
at constant speed $v_0$. Orientations and positions are updated following
(here in two dimensions):
\begin{eqnarray}
\theta_j^{t+1} &=& \arg \left[ \sum_{k\sim j}{\rm sign}[\cos(\theta_k^t -
\theta_j^t)] e^{i\theta_k^t} \right]+ \eta_j^t\\
{\bf r}_j^{t+1} &=& {\bf r}_j^t + v_0 \, {\bf e}(\theta_j^{t+1})
\end{eqnarray}
where ${\bf e}(\theta)$ is the unit vector along $\theta$, the sum is taken over particles $k$
within distance $d_0$ of particle $j$ (including $j$ itself),
and $\eta$ is a white noise
 with zero average and variance $\sigma^2$.
Like all Vicsek-style models,
it shows orientational order at large-enough global density
$\rho_0$ and/or small-enough noise
strength $\sigma$. It was shown in \cite{RODS} that the order is
nematic and that both the ordered
and disordered phases are subdivided in two: The homogeneous nematic phase
observed at low noise is replaced at larger $\sigma$ values by a segregated phase
where a dense, ordered band occupying a fraction of space coexists
with a disordered, dilute, gas. The transition
to disorder is given by the onset of a long-wavelength instability of this band observed
upon increasing $\sigma$ further. Finally, this chaotic regime
where dense ordered bands constantly form, elongate, meander, and disappear over very
long timescales is finally replaced by a ``microscopically-disordered'' phase at large noise
intensities.

Following \cite{BDG}, we write, in a dilute limit where only binary interactions
are considered and assuming that orientations are decorrelated between them
(``molecular chaos hypothesis''), a Boltzmann equation governing the evolution of the
one-particle distribution $f({\bf r},\theta,t)$:
\begin{equation}
\label{eqB}
\partial_t f({\bf r},\theta,t) + v_0\,{\bf e}(\theta) \cdot \nabla f({\bf r},\theta,t) =
I_{\rm dif}[f] +I_{\rm col}[f]
\end{equation}
with the angular diffusion and collision integrals
\begin{eqnarray}
I_{\rm dif}[f]\!\!\!&=&\!\! -\lambda f(\theta)+\!\!\lambda\int_{-\pi}^{\pi}\!\!\! d\theta' f(\theta')\!\!\int_{-\infty}^{\infty}\!\!\! d\eta\, P_\sigma(\eta)
\delta_{2\pi}(\theta'\!-\!\theta\!+\!\eta) \\
I_{\rm col}[f]\!\!\!&=&\!\!-f(\theta)\!\int_{-\pi}^{\pi} \!\!d\theta' K(\theta',\theta) f(\theta')
+\!\int_{-\pi}^{\pi}\!\!\! d\theta_1 f(\theta_1) \!\int_{-\pi}^{\pi}\!\!\!\! d\theta_2 \nonumber\\
\!&\times&\!\!
K(\theta_1,\theta_2) f(\theta_2)
\! \int_{-\infty}^{\infty} \!\!\!d\eta \, P_\sigma(\eta) \delta_{2\pi}(\Psi(\theta_1,\theta_2)\!-\!\theta\!+\!\eta)\nonumber
\end{eqnarray}
where $P_\sigma(\eta)$ is the microscopic noise distribution,
$\delta_{2\pi}$ is a generalized Dirac delta imposing
that the argument is equal to zero modulo $2\pi$,
$ K(\theta_1,\theta_2)=2 d_0 v_0 |{\bf e}(\theta_1)-{\bf e}(\theta_2)|$ is the
collision kernel for dilute gases \cite{BDG}, and
$\Psi(\theta_1,\theta_2)=\frac{1}{2}(\theta_1+\theta_2)+\frac{\pi}{2}(H[\cos(\theta_1-\theta_2)]-1)$ for $-\frac{\pi}{2}<\theta_2-\theta_1<\frac{3\pi}{2}$
(with $H(x)$ the Heaviside step function) codes for the nematic alignment.
Rescaling of time, space and density allows us to set the ``collision surface''
$S\equiv 2 d_0 v_0/\lambda =1$ and $v_0=1$ below, without loss of generality.

Next, the distribution function is expanded in Fourier series of the angle:
$f({\bf r},\theta,t) =\frac{1}{2\pi}\sum_{k=-\infty}^\infty f_k({\bf r},t) e^{-ik\theta}$,
with $f_k = f^*_{-k}$  and $|f_k|\leq f_0$.
The zero mode is nothing but the local density,
while $f_1$ and  $f_2$ give access to the polar and
nematic order parameter fields ${\bf P}$ and ${\bf Q}$:
\begin{equation}
\rho=f_0\,, \;
\rho {\bf P} \!=\! \left(\!\begin{array}{lr}
{\rm Re}f_1\\
{\rm Im}f_1 \end{array} \!\right) , \;
\rho {\bf Q} \!=\! \frac{1}{2} \!\left(\begin{array}{lr}
{\rm Re}f_2 & {\rm Im}f_2 \\
{\rm Im}f_2 & -{\rm Re}f_2\end{array} \!\right) \,.
\end{equation}
As a matter of fact, it is convenient to use $f_1$ and $f_2$, together with
the ``complex''operators $\triangledown \equiv \partial_x + i\partial_y$ and
$\triangledown^* \equiv \partial_x - i\partial_y$.
%
The continuity equation governing $\rho$ is given by integrating the Boltzmann
equation over angles:
\begin{equation}
\label{eqcont}
\partial_t \rho +  {\rm Re}(\triangledown^* f_1 ) \;=\; 0 \;.
\end{equation}
In Fourier space, the Boltzmann equation (\ref{eqB}) yields an infinite hierarchy
of equations:
\begin{eqnarray}
\!\!\!\partial_t f_k &\!\!+\!\!& \frac{1}{2}(\triangledown f_{k-1}\!+\!\triangledown^*\! f_{k+1})
= \nonumber\\
&&\!\!\!\! (\hat{P}_k-1) f_k
 + \frac{2}{\pi}\!\! \sum_{q=-\infty}^\infty\! \left[\hat{P}_k J_{kq} - \frac{4}{1\!-\!4q^2}\right]
f_q f_{k-q}
\label{eq:kgen}
\end{eqnarray}
where $\hat{P}_k=\int_{-\infty}^{\infty} d\eta P_{\sigma}(\eta) e^{i k \eta}$
and
\begin{eqnarray}
J_{kq} &=& \int_{-\frac{\pi}{2}}^{\frac{\pi}{2}} d\phi\, \left|\sin\frac{\phi}{2}\right|
\, e^{i(\frac{k}{2}-q)\phi}\\
\nonumber
&+& \cos \frac{k\pi}{2} \int_{\frac{\pi}{2}}^{\frac{3\pi}{2}}
d\phi\, \left|\sin\frac{\phi}{2}\right| \, e^{i(\frac{k}{2}-q)\phi}.
\end{eqnarray}

To truncate and close this hierarchy,
we adopt the following scaling structure, valid
near onset of {\it nematic} order, assuming,
in a Ginzburg-Landau-like approach \cite{IGOR-LEV},
small and slow variations of the density and of the polar and nematic fields:
\begin{equation}
\rho-\rho_0 \sim\epsilon, \;\;
\{f_{2k-1},f_{2k}\}_{k\ge 1} \sim \epsilon^k,
\; \;\triangledown  \sim \epsilon,\; \; \partial_t \sim \epsilon
\label{Ansatz}
\end{equation}
Note that the scaling of space and time is in line with the propagative
structure of our system, as seen in the continuity equation (\ref{eqcont}),
which contains no diffusion term.

\begin{figure}[t!]
\includegraphics[width=\columnwidth]{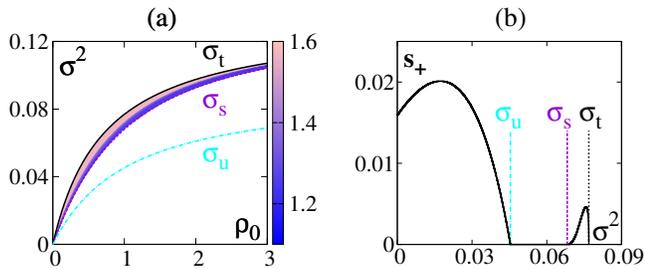}
\caption{(color online) (a) Linear stability of homogeneous solutions in the
$(\rho_0,\sigma)$ plane (plotted as a function of $\sigma^2$ to enhance clarity).
The line $\rho =\frac{15\pi(\hat{P}_{2}-1)}{40\hat{P}_{2}(2\sqrt{2}-1)-64}$,
given by $\mu=0$, is the basic instability line defining $\rho_{\rm t}(\sigma)$ or
$\sigma_{\rm t}(\rho_0)$: above it, the disordered homogeneous solution is linearly
stable; below, it becomes unstable and the ordered solution $f_2=\sqrt{\mu/\xi}$ exists.
This solution is unstable between the  $\sigma_{\rm t}$ and the $\sigma_{\rm s}$ lines.
It is linearly stable between $\sigma_{\rm s}$, and  $\sigma_{\rm u}$, which
marks the border of a region where $q=0$ is the most unstable mode.
The color scale codes for the angle between the most unstable wavevector and the
direction of nematic order.
(b) Largest eigenvalue $s_+$ (when positive) as a function of $\sigma^2$ for $\rho_0=1$.}
\label{fig1}
\end{figure}

The first non-trivial order yielding well-behaved equations is $\epsilon^3$:
keeping only terms up to this order, equations for $f_{k>4}$ identically vanish,
while those for  $f_3$ and $f_4$ provide expressions of these quantities
in terms of $\rho$, $f_1$, and  $f_2$, which allows us to write
the closed equations:
\begin{eqnarray}
\label{eqf1}
\partial_t f_1 &=& -\frac{1}{2} (\triangledown\rho + \triangledown^* f_2)
  + \frac{\gamma}{2} f^*_2 \triangledown f_2 \nonumber\\
&&-(\alpha - \beta|f_2|^2)f_1 + \zeta f^*_1 f_2\\
\label{eqf2}
\partial_t f_2 \!&=&\! -\frac{1}{2} \triangledown f_1 +\frac{\nu}{4} \triangledown\triangledown^* f_2
-\frac{\kappa}{2} f^*_1 \triangledown f_2 -\frac{\chi}{2} \triangledown^* (f_1 f_2)\nonumber\\
&&+(\mu-\xi |f_2|^2) f_2 +\omega f_1^2 +\tau |f_1|^2f_2
\end{eqnarray}
where all coefficients depend only on
the noise strength $\sigma$ (via the $\hat{P}_{k}$ coefficients) and the local density $\rho$:
\begin{equation}
\label{eqcoeff}
\begin{array}{lcl}
\hspace{-0.2cm}\nu =\left[\frac{136}{35\pi}\rho+1-\hat{P}_{3}\right]^{-1}&
\hspace{-0.2cm}& \omega =  \frac{8}{\pi}\left[\frac{1}{6}-\frac{\sqrt{2}-1}{2}\hat{P}_{2}\right] \\
\hspace{-0.2cm}\mu = \frac{8}{\pi}\left[\frac{2\sqrt{2}-1}{3}\hat{P}_{2}\!-\!\frac{7}{5}\right]\rho-1\!+\!\hat{P}_{2} &
\hspace{-0.2cm}& \zeta = \frac{8}{5\pi}\\
\hspace{-0.2cm}\alpha = \frac{8}{\pi} \left[\frac{1}{3}-\frac{1}{4}\hat{P}_{1}\right]\rho+1-\hat{P}_{1} &
\hspace{-0.2cm}& \chi = \nu\frac{2}{\pi} \left[\frac{4}{5}+\hat{P}_{3}\right]\\
\hspace{-0.2cm}\kappa = \nu \frac{8}{15}\left[\frac{19}{7}-\frac{\sqrt{2}+1}{\pi}\hat{P}_{2}\right] &
\hspace{-0.2cm}&\gamma = \nu\frac{4}{3\pi} \left[\hat{P}_{1}-\frac{2}{7}\right]  \\
\hspace{-0.2cm}\tau = \chi \frac{8}{15}\left[\frac{19}{7}-\frac{\sqrt{2}+1}{\pi}\hat{P}_{2}\right] &
\hspace{-0.2cm}& \beta = \gamma\frac{2}{\pi}\left[\frac{4}{5}+\hat{P}_{3}\right] \\
\multicolumn{3}{l}{\hspace{-0.2cm}\xi = \frac{32}{35\pi}\!
\left[\frac{1}{15}\!+\!\hat{P}_{4}\right]\!
\left[\frac{13}{9}\!-\!\frac{6\sqrt{2}+1}{\pi}\hat{P}_{2}\right]\!
\left[\frac{8}{3\pi}\!\left(\frac{31}{21}\!+\!\frac{\hat{P}_{4}}{5}\right)\rho+1\!-\!\hat{P}_{4}\right]^{-1}
}
\end{array}
\end{equation}
Below, we specialize for convenience to the Gaussian noise distribution
$P_\sigma(\eta)=\frac{1}{\sigma\sqrt{2\pi}} \exp[-\frac{\eta^2}{2\sigma^2}]$ for which
 $\hat{P}_k=\exp[-\frac{1}{2} k^2\sigma^2]$ \cite{NOTE4}.
A few remarks are then in order.
First, $\mu$ can change sign and become positive for large enough $\rho$,
while $\alpha$ is always positive: The homogeneous disordered
state ($f_1\!=\!f_2\!=\!0$) undergoes an instability to nematic order when $\mu=0$,
defining the basic transition line $\sigma_t(\rho_0)$ in the $(\rho_0,\sigma)$ plane
(Fig.~\ref{fig1}a).
Next, $\xi$ being positive
in the $\mu>0$ region where the disordered solution is unstable,
Eqs.~(\ref{eqf1}-\ref{eqf2}) possess a homogeneous nematically-ordered solution
$(f_1,f_2)=(0,\sqrt{\mu/\xi})$ (assuming the order along $\mathbf{x}$, so that $f_2$
is real positive) \cite{NOTE}.

In the following, we restrict ourselves to slightly modified coefficients by suppressing the
local density dependence (replacing $\rho$ by $\rho_0$) of $\nu$ and $\xi$,
and thus of all other coefficients but the crucial linear ones $\mu$ and
$\alpha$
(note that this corresponds to retaining only terms up to order $\epsilon^3$
after expanding coefficients in $\delta\rho$). This expansion does not change
any of the main features of our equations, but allows us to find exact front solutions (see below).

We have studied the linear stability of the homogeneous nematic solution with respect to
perturbations of arbitrary wavevector in the full $(\rho_0,\sigma)$
parameter plane (Fig.~\ref{fig1}). Similarly to the polar case with
ferromagnetic alignment \cite{BDG,IHLE,MCM-polar,MCM-polar2},
this solution is unstable to long wavelengths in
a region bordering the basic transition line. The most unstable modes in this region are
roughly ---but not exactly--- transversal to the order of the solution \cite{NOTE6}.
The homogeneous nematic solution
becomes linearly stable deeper in the ordered phase
(line $\sigma_{\rm s}$ in Fig.~\ref{fig1}a), but its stability domain is limited
by another instability region where $q=0$ is the most unstable mode
(line $\sigma_{\rm u}$, which can be shown to be given by
 $\alpha+\zeta \sqrt{\mu/\xi}-\beta \mu/\xi=0$).
This strong instability, which occurs at large
densities and/or weak noise, may be an artifact introduced by our truncation \cite{NOTE3}.

\begin{figure}[t!]
\includegraphics[width=0.45\columnwidth]{fig2a-new.eps}
\includegraphics[width=0.45\columnwidth]{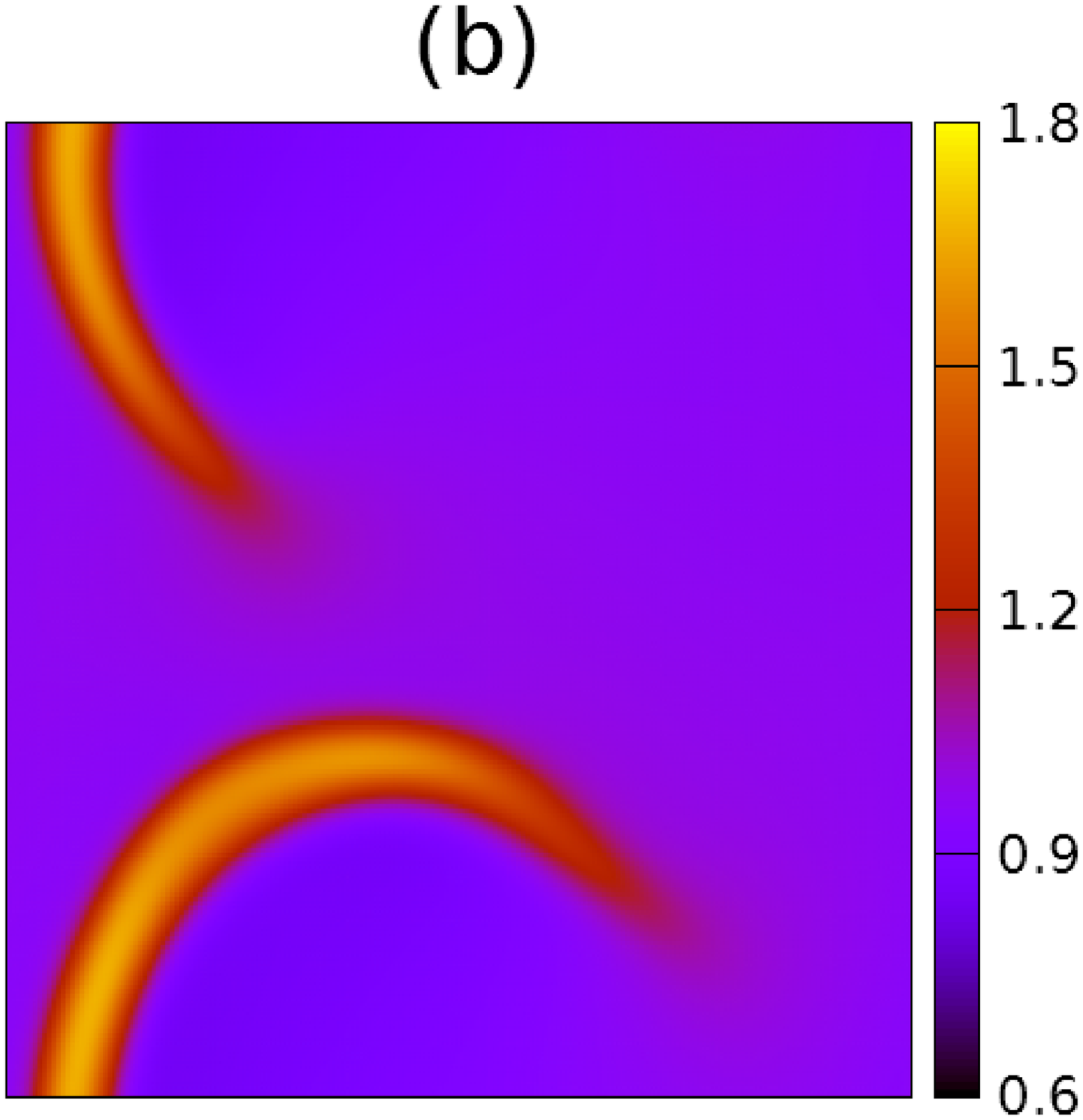}
\vspace{0.5cm}
\includegraphics[width=0.45\columnwidth]{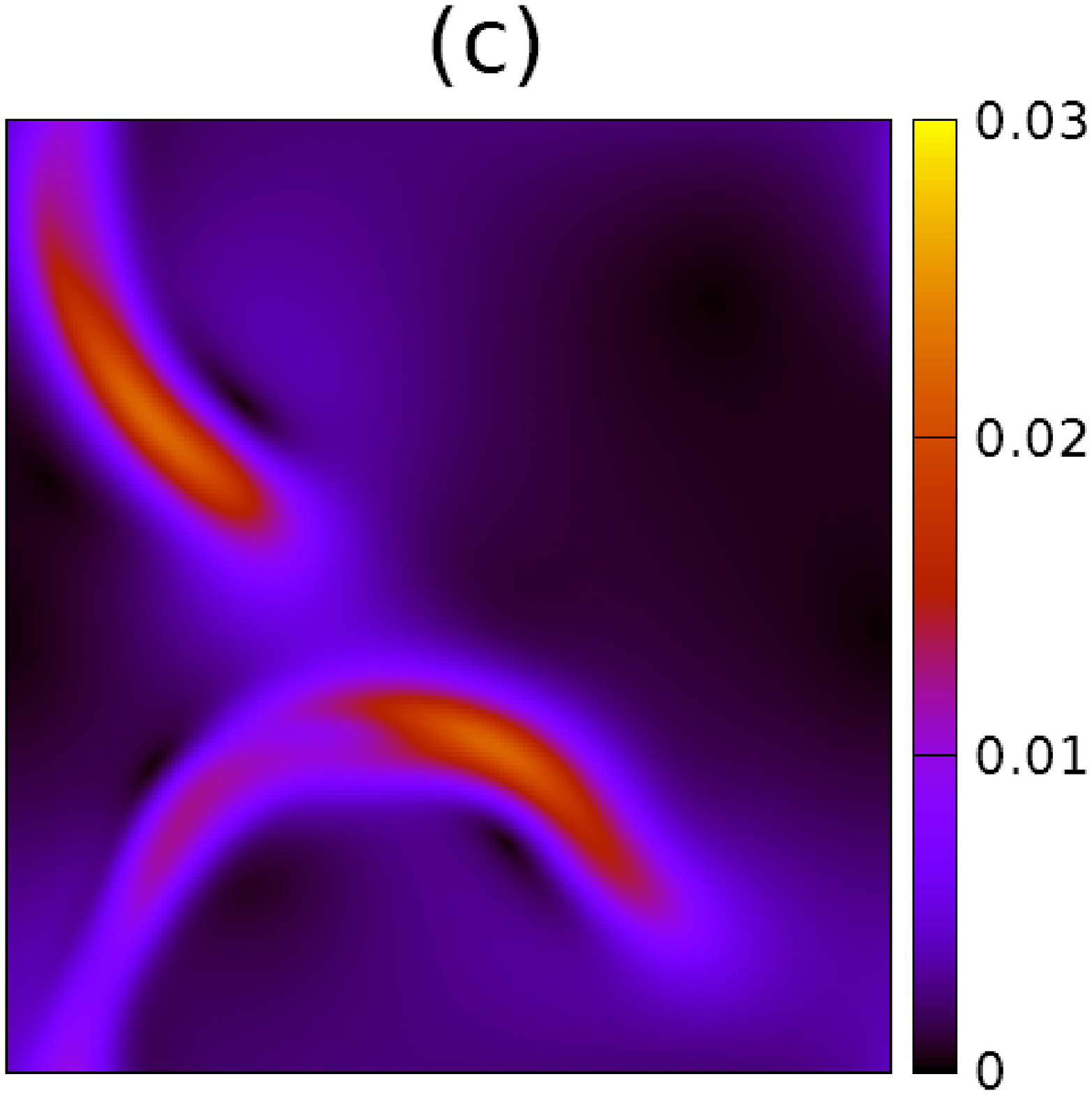}
\includegraphics[width=0.45\columnwidth]{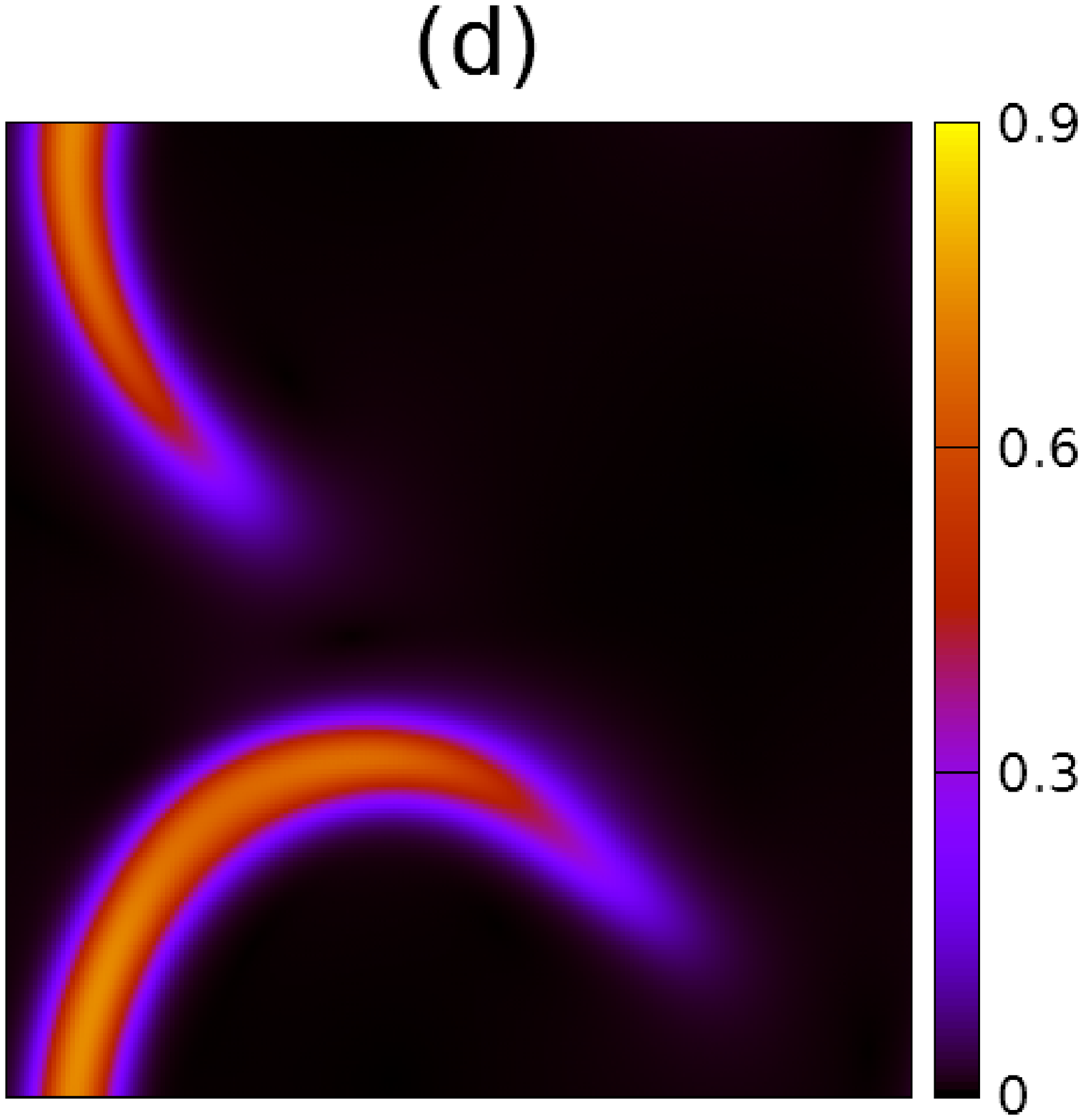}
\vspace{-0.5cm}

\caption{(color online) Numerically obtained density-segregated solutions.
(a) density and $f_2$ profiles of a stationary banded solution
($f_1=0$ throughout). The fronts linking the disordered and ordered domains
can be perfectly fitted to hyperbolic tangents (not shown).($\sigma=0.26,\rho_{0}=1,L=200$)
 (b,d,e) chaotic band regime: snapshots of (respectively) $\rho$,  $|f_1|$, and $|f_2|$. ($\sigma=0.2826,\rho_{0}=1,L=200$)}
\label{fig2}
\end{figure}

To go beyond the linear stability analysis of spatially homogeneous solutions,
we performed numerical integrations of Eqs.~(\ref{eqf1},\ref{eqf2})
in rectangular domains with periodic boundary conditions of typical linear sizes 50-200 \cite{CODE}.
For parameter values in the region of transversal instability of the nematic
homogeneous solution,
we observe, after transients and from almost every initial condition,
stationary asymptotic solutions in which nematic order is confined to
and oriented along a dense band with local density $\rho_{\rm band}$ amidst a homogeneous
disordered ``gas'' with $\rho_{\rm gas}$ such that
$\rho_{\rm band}>\rho_{\rm s}>\rho_{\rm t}>\rho_{\rm gas}$ (Fig.~\ref{fig2}a),
where $\rho_{\rm s}(\sigma)$ is given by inverting $\sigma_{\rm s}(\rho_0)$.
Varying system size and using various domain aspect ratios, we find most often
a single band oriented along the shortest dimension of
the domain, which occupies a size-independent fraction $\Omega$ of space.
All these observations are in agreement with the behavior of the original microscopic
model \cite{RODS}.

\begin{figure}[t!]
\includegraphics[width=\columnwidth]{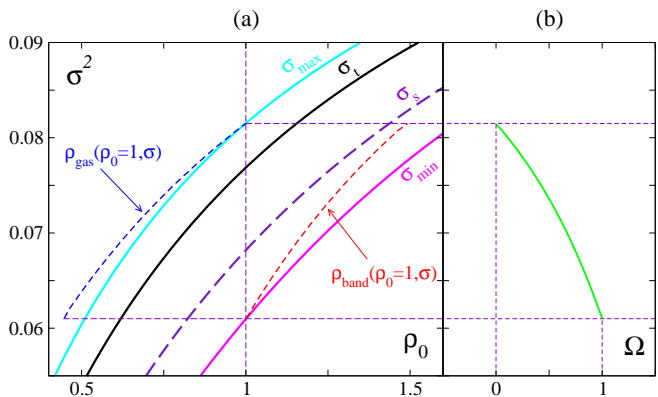}
\caption{(color online) Analytic band solutions for the slightly simplified
system (see text). (a) $(\rho_0,\sigma)$ parameter plane with basic instability
line $\sigma_{\rm t}$, stability limit of homogeneous ordered phase $\sigma_{\rm s}$,
and limits of existence of band solutions  $\sigma_{\rm min}$ and  $\sigma_{\rm max}$.
The short-dashed blue and red lines show the $\rho_{\rm gas}$ and $\rho_{\rm band}$
density values of the band solutions for $\rho_0=1$ as a function of $\sigma$ 
over their existence range $[\sigma_{\rm min},\sigma_{\rm max}]$,
indicated by the thin horizontal dashed violet lines.
(b) variation with $\sigma$ of $\Omega$,
the fraction of space occupied by the ordered part of the band solution, for  $\rho_0=1$.
}
\label{fig3}
\end{figure}

Band solutions are also present {\it beyond} the parameter domain
where the homogeneous ordered state is linearly unstable.
Starting from such a solution found at other parameter values ---or from
sufficiently inhomogeneous initial conditions---, we find band solutions
both for $\rho_0$ values larger than $\rho_{\rm s}$,
where they coexist with the homogeneous ordered phase, and below  $\rho_{\rm t}$,
where the disordered homogeneous solution is linearly stable.
Working at fixed $\rho_0$ varying $\sigma$ for clarity, we thus find
bands in a $[\sigma_{\rm min},\sigma_{\rm max}]$ interval larger than the
linear-instability interval $[\sigma_{\rm s},\sigma_{\rm t}]$ (Fig.~\ref{fig3}).
Along it, the fraction occupied by the ordered
band decreases from  $\Omega\lesssim 1$ near $\sigma_{\rm min}$
to  $\Omega\gtrsim 0$ near $\sigma_{\rm max}$.
Furthermore, within a small layer $\sigma_{\rm max}<\sigma\le\sigma_{\rm c}$,
the bands are unstable, giving rise to a chaotic
regime where they twist, elongate, break, and form again,
in a manner strikingly similar to observations made in the original microscopic model
(Fig.~\ref{fig2}b, \cite{EPAPS}).

Thus the region of linear instability of the homogeneous ordered solution
does {\it not} correspond to the existence (and stability) domain of band
solutions, which is wider. In the original microscopic model, with its built-in
fluctuations, coexistence of band solutions and homogeneous order has not
been reported, but the homogeneous solution was found metastable near the
threshold of emergence of bands where these appear ``suddenly'' \cite{RODS}.
At the other end of the band existence region, no coexistence was reported
between band solutions (chaotic or not)
and the homogeneous disordered state,
suggesting that the latter is always driven to the former by intrinsic fluctuations.
Although this calls for revisiting the microscopic model,
this suggests that its thresholds do {\it not} correspond to those given by
the linear stability of the homogeneous solutions of our continuous equations.

We now show that many of the features of band solutions can be found
analytically.
Suppose that, as observed, $f_1=0$ for band solutions and
that $f_2$ is real and positive (i.e., nematic order is along $x$),
and depends only on $y$.
For a stationary solution, Eq.~(\ref{eqf1}) then yields, after integration over $y$,
\begin{equation}
\label{rhos}
\rho - f_2 - \frac{1}{2}\gamma f_2^2=\tilde{\rho}
\end{equation}
where $\tilde{\rho}$ is a constant. This allows us to write Eq.~(\ref{eqf2}), again looking
for stationary solutions, in terms of $f_2$ only:
\begin{equation}
\label{eqf2bis}
\frac{\nu}{4}\partial_{yy}f_2=-\mu' f_2(\tilde{\rho}-\rho_{\rm t}+f_2)+\left[\xi-\frac{\gamma}{2}\mu'\right]f_2^3
\end{equation}
where we have rewritten  $\mu=\mu'(\rho-\rho_{\rm t})$, with $\mu'$
independent of $\rho$.
Eq.~(\ref{eqf2bis}) can be interpreted as an ordinary differential
equation describing the motion in fictitious time $y$ of a particle of mass
$\frac{\nu}{4}$ in a potential with one maximum at $f_2=0$ and another at some value
$\tilde{f}_2$ (close enough to threshold).
A band solution is found for a trajectory starting at and returning to the same fixed point $f_2=0$ (homoclinic orbit).
Direct integration of  Eq. (\ref{eqf2bis}) yields, under the condition $\lim_{y\rightarrow\pm\infty}f_{2}(y)=0$, the following solution
\begin{equation}
f_2(y)= \frac{3 (\rho_t-\tilde \rho )}{1+a \cosh
   \left(2 y \sqrt{\mu' (\rho_{\rm t}-\tilde\rho )/\nu }\right)}
   \label{band}
\end{equation}
where $a=\sqrt{1+9b(\tilde{\rho}-\rho_t)/2\mu'}$ and $b=\xi-\mu'\gamma/2$.
The value of $\tilde{\rho}$ can be obtained from the condition $\int_{L}\rho\left(y\right)dy=L\rho_{0}$, 
where \emph{L} is the size of the box. We can neglect the exponentially decreasing tails in the integral
and solve the equation
$\int_{-\infty}^{\infty}\left(\rho\left(y\right)-\tilde{\rho}\right)dy=L\left(\rho_{0}-\tilde{\rho}\right)$.
Under the assumption $L\rightarrow\infty$ we obtain
\begin{equation}
\tilde{\rho}\approx\rho_{t}-\frac{2\mu'}{9b}\left(1-K_{1}e^{-K_{2}L}\right)
\label{tilde_rho}
\end{equation}
where $K_{1}$ and $K_{2}$ are positive quantities depending on $\sigma$ and $\rho_0$
whose expression we omit
for compactness. Substituting this value in the expression of \emph{a} gives
us $a=K_{1}e^{-K_{2}L/2}$, yielding a width of the band proportional to $L$, in agreement with
observations on the microscopic model.
As $L\rightarrow\infty$, the value of $\tilde{\rho}$
converges to the asymptotic value $\tilde{\rho}_{\rm gas}$.

To determine the surface fraction $\Omega$ occupied by the ordered band,
we use the relation
$\Omega\left(\rho_{\rm band}-\rho_{\rm gas}\right)+\rho_{\rm gas}=\rho_0$.
Substituting the value of $\rho_{\rm band}$ obtained from
Eqs.~(\ref{rhos}) and (\ref{band}) at $y=0$, we find for $L\rightarrow\infty$
\begin{equation}
\Omega=\frac{9b^{2}\left(\rho_{0}-\rho_{t}\right)+2b\mu'}{2\mu'\left(\gamma\mu'+3b\right)}\;.
\end{equation}
The condition $0<\Omega<1$ yields the lower limit $\sigma_{\rm min}$
and the upper limit $\sigma_{\rm max}$ of existence of the band solution.
All these results are presented in Fig.~\ref{fig3}.

To summarize, using a  ``Boltzmann-Ginzburg-Landau'' controlled expansion scheme,
we derived a set of minimal yet complete nonlinear field equations from
the Vicsek model with nematic alignment studied in \cite{RODS}. This simple setting
allowed for a comprehensive analysis of the linear and nonlinear dynamics
of the field equations obtained because our approach automatically yields
a ``meaningful manifold'' parameterized by global density and noise strength in
the high-dimensional space spanned by all coefficients of the continuous equations.
Excellent semi-quantitative agreement was found with the simulations of the
original microscopic model.
An analytic picture of the nonlinear banded solutions present was
obtained. Their existence domain was found different from the region of
linear instability of the homogeneous ordered phase, stressing the importance of
a nonlinear analysis. More work, beyond the scope of this Letter,
is needed to obtain a better understanding of
the chaotic regimes observed.  To this aim, we plan to study the linear
stability of the band solutions in two dimensions.

Our equations (\ref{eqf1}-\ref{eqf2}) are simpler than those written
by Baskaran and Marchetti in \cite{MCM-rod1,MCM-rod2},
not only because our microscopic starting point does not include positional diffusion of
particles. The method used there seems intrinsically different, yielding
more terms, many with a different structure from ours, while some of our nonlinear ones
do not appear. It is not known yet whether the equations of \cite{MCM-rod2} can also account
for the nonlinear phenomena described above. Future work should explore this point
in some detail.

Finally, recent experiments have revealed that microtubules
displaced by a ``carpet'' of dynein motors grafted to a substrate self-organize
their collective motion into large vortical structures \cite{Sumino2012}.
It was shown that
a Vicsek model with nematic alignment, but one in which the microscopic
noise is colored, accounts quantitatively
for the observed phenomena. Extending the approach followed here to this case is the subject
of ongoing work.

\acknowledgements
We thank the Max Planck Institute for the Physics of Complex Systems, Dresden,
for providing the framework of the Advanced Study Group
``Statistical Physics of Collective Motion''
within which much of this work was conducted. 
The work of I.S.A. was supported by the U.S. Department of Energy, Office of Basic Energy Sciences, Division of Materials Science and Engineering, under Contract DEAC02-06CH11357.


\begin{thebibliography}{99}
\bibitem{NOTE4} For Gaussian noise, and other unimodal,
centered noise distributions, most of the
coefficients above do not change sign and are positive in the
parameter region of interest.
\bibitem{NOTE} Other homogeneous solutions with both $f_1$ and $f_2$ non zero exist,
but they are not observed in simulations, being either unphysical (having a  modulus
larger then $f_0$) and/or unstable.
\bibitem{NOTE6} {This near-threshold instability is generically present in all dry
active matter systems with metric interactions. See, e.g.,
\cite{TOPOKINETIC,MCM-polar2}.}
\bibitem{CODE} 
 We used a pseudospectral code with Euler
  timestepping ($\Delta t = 0.01$), with at least $400\times 400$ Fourier
  modes on square domains of linear size $L=200$.
\bibitem{NOTE3} Simulations of Eqs.~(\ref{eqcont},\ref{eqf1},\ref{eqf2})
in this region lead to unbounded growth, whereas the original
Vicsek model with nematic alignment shows no sign of instability.
Discarded higher-order terms may be needed to stabilize our equations away
from the onset of nematic order.
\end{thebibliography}
\end{document}